\documentclass[aps,twocolumn]{revtex4}
\usepackage{epsfig}
\usepackage{graphicx}

\begin{document}
\pagestyle{plain}

\title{Features and dimensions: Motion estimation in fly vision}

\author{William Bialek$^a$ and Rob R. de Ruyter van Steveninck$^b$}

\affiliation{$^a$Joseph Henry Laboratories of Physics, $^b$Department of Molecular Biology, and the Lewis--Sigler Institute for Integrative Genomics\\
Princeton University, Princeton, New Jersey 08544\\
$^b$Department of Physics, Indiana University, Bloomington, Indiana 47405}

\begin{abstract}
We characterize the computation of motion in the fly visual system as a
mapping from the high dimensional space of signals in the retinal
photodetector array to the probability of generating an action potential
in a motion sensitive neuron.  Our approach to this problem identifies a low
dimensional subspace of signals within which the neuron is most
sensitive, and then samples this subspace to visualize the nonlinear
structure of the mapping.  The results illustrate the  computational
strategies predicted for a system that
makes optimal motion estimates given the physical noise sources in the
detector array.  More generally, the hypothesis that neurons are sensitive to
low dimensional subspaces of their inputs formalizes the intuitive notion of
feature selectivity and suggests a strategy for characterizing the neural
processing of complex, naturalistic sensory inputs.
\end{abstract}

\date{\today}

\maketitle

\section{Introduction}

Vision begins with the counting of photons by a large array of detector elements in the retina. 
From these inputs, the brain is thought to extract features, such as the
edges in an image or the velocity of motion across the visual field, out of which our perceptions are constructed.  In some cases    we can point to individual neurons in the brain that represent
the output of this feature extraction; classic examples include the
center--surround comparison encoded by retinal ganglion cells \cite{barlow_53},  the orientation selective ``edge detectors'' described by Hubel
and Wiesel  in primary visual cortex \cite{hubel+wiesel_62},  and the direction
selective, motion sensitive neurons found in visual systems from flies \cite{bishop+keehn_66} to rabbits \cite{barlow+levick_65}
to primates \cite{baker+al_81}.  As
emphasized by Marr, feature extraction seems such a natural and
elementary step in sensory signal processing that it is easy to overlook
the challenges posed by these computations \cite{marr_82}. 

Our focus in this paper is on the computations leading to the extraction
of motion across the visual field, although we believe that  the key
issues are common to many different  problems in neural
computation.    In primates the spike trains of motion sensitive neurons are correlated with
perceptual decision making on a trial--by--trial basis \cite{britten+al_96}, lesions to populations of these neurons produce specific
behavioral deficits \cite{newsome+al_85},
and stimulation of the population can bias both
perceptual decisions \cite{salzman+al_90} and more graded visuomotor behaviors
\cite{groh+al_97}.   In insects, ablation of single motion sensitive
neurons leads to deficits of visuomotor behavior that match the spatial
and directional tuning of the individual neurons \cite{hausen+wehrhahn_83},  and  one can use the 
spike sequences from these cells to estimate  the trajectory of motion
across the visual field or to distininguish among subtly different
trajectories \cite{rieke+al_97}.  Strikingly, at least under some conditions the precision of these
motion estimates approaches the physical limits set by diffraction and
photon shot noise at the visual input \cite{bialek+al_91, ruyter+bialek_95, note1}.  Taken together these observations suggest strongly that the
motion sensitive neurons represent most of what the organism knows about visual
motion, and in some cases everything that the organism could know given the
physical signals and noise in the retina.

It is tempting to think that the stimulus for a motion sensitive neuron
is the velocity of motion across the visual field, but this is wrong:  the
input to all visual computation is a representation of the spatiotemporal history of
light intensity falling on the retina, $I({\bf\vec x},t)$. This representation is approximate, first because the physical carrier, a photon stream, is inherently noisy, and second because the intensity pattern is blurred by the optics, and sampled in a discrete raster. Features, such as velocity, must be computed explicitly from this raw input stream. 
As discussed below, even the simplest visual computations have
access to $D \sim 10^2$ spacetime samples of  
$I({\bf\vec x},t)$.
If the response of a single neuron were an
arbitrary function on a space of 100 dimensions, then no reasonable
experiment would be sufficient to characterize the computation that is
represented by the neuron's output spike train.  Any method for
characterizing the mapping from the visual input $I({\bf\vec x},t)$ to an
estimate of motion as encoded by a motion sensitive neuron must thus
involve some simplifying assumptions. 

Models for the neural computation of motion go back
(at least) to the classic work of Hassenstein and Reichardt,
who proposed that insects compute motion by evaluating a
spatiotemporal correlation of the signals from the array of photodetector
cells in the compound eye \cite{hassenstein+reichardt_56}.  Essentially the
same computational strategy is at the core of the motion energy
models  that are widely applied to the analysis
of human perception and neural responses in primates \cite{adelson+bergen_85}.  Both the
correlation model and the motion energy model have been extended in 
various ways to include saturation or normalization  
of the responses 
\cite{heeger_92, simoncelli+heeger_98}.  A seemingly very different approach emphasizes that
motion is a relationship between spatial and temporal variation in the
image, and in the simplest case this means that velocity should be
recoverable as the ratio of temporal and spatial derivatives 
\cite{limb+murphy_75}.
Finally, the fact that the fly visual system
achieves motion estimates with  a precision close to the physical limits \cite{bialek+al_91, ruyter+bialek_95}  motivates the
theoretical question of which   estimation strategies will in fact make best
use of the available signals, and this leads to rather specific predictions
about the form of the motion computation \cite{potters+bialek_94}.  The work described here has its origins in the attempt to test these
predictions of optimal estimation theory.

The traditional approach to testing theories of motion estimation
involves the design of particular visual stimuli which would highlight or
contrast the predictions of particular models.  This tradition is best
developed in the work of the Reichardt school, which aimed at testing and
elaborating the correlation model for motion estimation in fly vision
\cite{reichardt+poggio_76, poggio+reichardt_76}.  The fact that
simple mathematical models developed from elegant behavioral experiments
in the early 1950s provided a basis for the design and analysis of
experiments on the responses of single neurons decades later \cite{single+borst_98} should be
viewed as one of the great triumphs of theoretical approaches to brain
function.   While the correlation model (or the related motion energy models)
describes many aspects of the neural response, it probably is fair to say that
the simplest versions of these models are not sufficient for describing neural responses to motion generally, and especially in more natural conditions.

One of the clear predictions of optimal estimation theory  is that computational strategies which make the best use of
the available information in the retinal array must adapt to take account
of different stimulus ensembles \cite{potters+bialek_94}:  not only will the
computation of motion reach a different answer in (for example) a
bright, high contrast environment and in a dim, low contrast environment,
the optimal strategy actually involves {\em computing a different
function} in each of these environments.  Further, this adaptation in
computational strategy is not like the familiar light and dark
adaptation; instead it involves adjusting the brain's computational strategy in
relation to the whole {\em distribution} of visual inputs rather than
just the mean.   If statistical adaptation occurs, then the program of
testing computational models by careful choice of stimuli has  a major
difficulty, namely that the system will adapt to our chosen stimulus
ensemble and we may not be able to isolate different aspects  of the
computation as expected.  

There is evidence that the coding of
dynamic signals adapts to the input distribution both in the fly motion
sensitive neurons \cite{ruyter+al_96, brenner+al_00, fairhall+al_01}
and in the vertebrate retina \cite{smirnakis+al_97, meister+berry_99}, so that in these systems at least the representation of stimulus features depends on the context in
which they are presented.  In these examples context is defined by the probability distribution from which the signals are drawn, but there also is a large body of work demonstrating that neural responses at many levels of the visual system are modulated by the (instantaneous) spatial context in which localized stimuli are presented \cite{allman+al_85}.  What is needed, then, is a method which allows us to
analyze the responses to more complex---and ultimately to fully
natural---inputs and decompose these responses into the elementary
computational steps.  

We will see that models of
motion estimation share a common structure:  estimation involves a
projection of the high dimensional visual inputs onto a lower dimensional
space, followed by a nonlinear interaction among variables in this low
dimensional space.  The main goal of this paper, then,  is to present an
analysis method that allows us to observe directly the small number of
dimensions which are relevant to the processing of complex, high dimensional
inputs by any particular neuron.   We use this approach to show that
the motion sensitive neuron H1 in the fly visual system computes a
function of its inputs that is of the form predicted by optimal
estimation theory.  More generally, we believe that the reduction of
dimensionality may be the essential simplification required for progress on
the subjects of neural coding and processing of complex, naturalistic sensory signals.

\section{Models of motion estimation}
\label{models}

\begin{figure}[b]
\begin{center}
\epsfxsize=\linewidth
\epsfbox{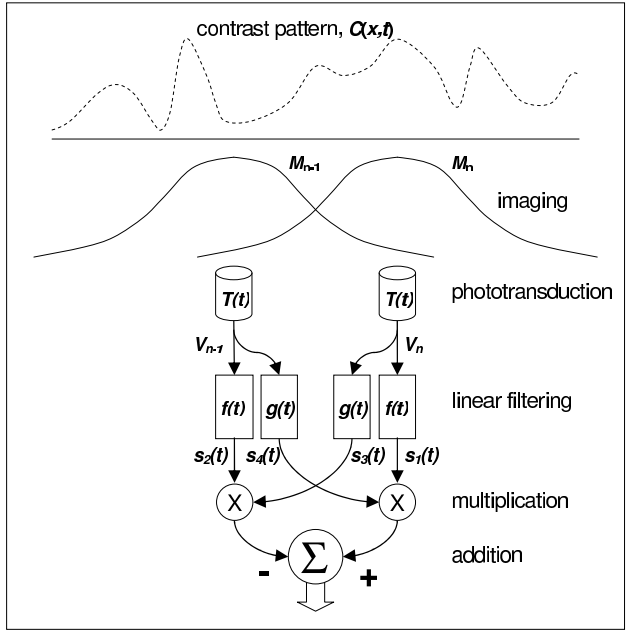}
\caption{The correlator model of visual motion detection, adapted from Ref. \cite{hassenstein+reichardt_56}. A spatiotemporal contrast pattern $C(x,t)$ is blurred by the photoreceptor point spread function, $M(x)$, and sampled by an array of photoreceptors, two of which (neighboring photoreceptors numbers $n-1$ and $n$) are shown here. After phototransduction, the signals in each photoreceptor are filtered by two different linear filters, $f(t)$ and $g(t)$. The outputs of these filters from the different photoreceptors, $s_1(t)$ and $s_3(t)$ from photoreceptor $n$ and $s_2(t)$ and $s_4(t)$ from photoreceptor $n-1$ are multiplied and one of these products is subtracted from the other by the addition unit, yielding a direction selective response.}
\label{reichardt_fig}
\end{center}
\end{figure}

The classic model of visual motion detection is the Reichardt correlator,
schematized in Fig. \ref{reichardt_fig}.  In the simplest version of the model, the output
signal is just the product of the voltages in neighboring
photoreceptors that have been passed through different filters,
\begin{eqnarray}
{\dot \theta}_{\rm est} (t) &\approx& \left[ \int d\tau f(\tau) V_{\rm n}(t
-\tau)\right] \nonumber\\
&&\,\,\,\,\,\,\,\,\,\,\times \left[
\int d\tau' g(\tau ') V_{\rm n-1} (t-\tau')\right] .
\end{eqnarray}
This signal has a directionality, since the ${\rm n}^{th}$ receptor voltage is
passed through filter $f$ while its left neighbor is passed through filter
$g$.  A better estimate of motion, however, would be genuinely antisymmetric
rather than merely directional, and to achieve this we can subtract the signal
computed with the opposite directionality:
\begin{eqnarray}
{\dot \theta}_{\rm est} (t) &=&  \left[
\int d\tau f(\tau) V_{\rm n}(t -\tau) \right] 
\nonumber\\
&&\,\,\,\,\,\,\,\,\,\,\times \left[
\int d\tau' g(\tau') V_{\rm n-1} (t-\tau') \right]
\nonumber\\
&&-
\left[ \int d\tau f(\tau) V_{\rm n-1}(t -\tau) \right] \nonumber\\
&&\,\,\,\,\,\,\,\,\,\,\,\,\,\,\,\times \left[
\int d\tau' g(\tau ') V_{\rm n} (t-\tau')\right].
\label{corr2}
\end{eqnarray}

Although it is natural to discuss visual computation using the photoreceptor
signals as input, in fact we can't control these signals and so we should
refer the computation back to image intensities or contrasts.  If the
photoreceptors give linear responses to contrast $C({\bf\vec x},t)$ over some
reasonable range, then we can write
\begin{equation}
V_{\rm n} (t) = \int d^2 x M({\bf\vec x}-{\bf\vec x}_{\rm n}) \int
d\tau T(\tau )C({\bf\vec x},t-\tau) ,
\label{VfromC}
\end{equation}
where $M({\bf\vec x})$ is the spatial transfer function or aperture of the
receptor and $T(\tau)$ is the temporal impulse response.  Substituting into
Eq. (\ref{corr2}), we obtain
\begin{equation}
{\dot \theta}_{\rm est}^{(n)} (t) = s_1(t) s_4(t) - s_2(t) s_3(t) ,
\label{fourpieces}
\end{equation}
where we are now careful to note that this is the estimate obtained at point $n$ on the retina, and
each of the signals $s_i(t)$ is a linearly filtered version of the
spatiotemporal contrast pattern,
\begin{eqnarray}
s_1(t) &=& \int d^2 x d\tau M({\bf\vec x} - {\bf\vec x}_{\rm n}) 
{\hat f}(\tau) C({\bf\vec x}, t-\tau )
\label{s_1}\\
s_2(t) &=& \int d^2 x d\tau M({\bf\vec x} - {\bf\vec x}_{\rm n-1}) 
 {\hat f}(\tau) C({\bf\vec x}, t-\tau )
\\
s_3(t) &=& \int d^2 x d\tau M({\bf\vec x} - {\bf\vec x}_{\rm n}) 
 {\hat g}(\tau) C({\bf\vec x}, t-\tau ) , 
\\
s_4(t) &=&  \int d^2 x d\tau M({\bf\vec x} - {\bf\vec x}_{\rm n-1}) 
 {\hat g}(\tau) C({\bf\vec x}, t-\tau )
\label{s_4}
\end{eqnarray}
with the temporal filters
\begin{eqnarray}
{\hat f}(\tau) &=& \int d\tau' f(\tau - \tau') T(\tau '),\\
{\hat g}(\tau) &=& \int d\tau' g(\tau - \tau') T(\tau ') .
\label{filterdefs}
\end{eqnarray}
Although much of the effort in applying the Reichardt
(and related) models to experiment has focused on measuring the particular
filters $f$ and $g$, we want to emphasize here the fact the {\em form} of Eq.
(\ref{fourpieces}) is simple no matter how complex the filters might be.

In principle, the estimate of motion at time $t$ can be influenced by the
entire movie that we have seen up to this point in time, that is by the whole
history $C({\bf\vec x},t-\tau)$ with $\tau >0$. In real systems it typically makes sense to use a finite time window, and in the concrete example of the fly's
motion sensitive neurons, the relevant window for the computation can be on the
order of 150 msec \cite{ruyter+al_86}, while the absolute time
resolution is at worst a few milliseconds \cite{ruyter+al_01}.   This means that there are at least 
$\sim 50$  time points which can enter our description of this movie.  To compute
motion we need access to at least two independent spatial pixels, so altogether the
history $C({\bf\vec x},t-\tau)$ involves at least one hundred numbers:  ``the
stimulus'' is a point in a space of over 100 dimensions.  Despite this
complexity of the stimulus, Eq. (\ref{fourpieces}) tells us that---if this
model of motion computation is correct---only four stimulus parameters are
relevant.  The computation of motion involves a nonlinear combination
of these parameters, but the parameters themselves are just linear
combinations of the $\sim 10^2$ stimulus variables.   While our immediate concern
is with the fly visual system, similar dimensionalities arise if we think about
the stimuli which drive neurons in the early stages of mammalian visual
cortex, or the acoustic stimuli of relevance for early stages of auditory processing.

More formally, we can  think of the stimulus history $C({\bf\vec x},t-\tau)$
leading up to time $t$ as a vector ${\bf \vec s}_t$ in $D\sim 10^2$ dimensions.  If
the motion sensitive neurons encode the output of the simplest Reichardt
correlator then the probability per unit time $r(t)$ of generating an action
potential will be of the form
\begin{eqnarray}
r(t) &=& {\bar r}  G(s_1, s_2, \cdots s_K)\\
\label{modeldef}
s_1 &=& {\bf {\hat v}}_{\rm 1}{\bf \cdot \vec s}_t\\
s_2 &=& {\bf {\hat v}}_{\rm 2}{\bf \cdot \vec s}_t \\
&\cdots , &
\end{eqnarray}
where in this case $K=4$ and the vectors ${\bf {\hat v}}_{\rm i}$ describe the
spatiotemporal filters from Eq's. (\ref{s_1}--\ref{s_4}).  
The central point is {\em not} that the
function $G$ has a simple form---it might not, especially when we consider the
nonlinearities associated with spike generation itself---but rather that the
number of relevant dimensions $K$ is much less than the full stimulus dimensionality
$D$.

As described here, the correlation computation involves just two photoreceptor elements.  In motion energy models  these individual detector elements are replaced by potentially more complex spatial receptive fields \cite{adelson+bergen_85}, so that $M({\bf\vec x})$ is Eq. (\ref{VfromC}) can have a richer structure than that determined by photoreceptor optics; more generally we can imagine that rather than two identical but displaced spatial receptive fields we just have two different fields.  The general structure is  the same, however:  two spatial samples of the movie $C({\bf\vec x},t)$ are passed through two different temporal filters, and the resulting four variables are combined is some appropriate nonlinear fashion.  Elaborated versions of both the Reichardt and motion energy models might include six or eight projections, but the space of relevant stimulus variables always is much smaller than the hundreds of dimensions describing the input stimulus movie.

Wide field motion sensitive neurons, such as the fly's H1 cell which is the subject of the experiments below, are thought to sum the outputs of many elementary pairwise correlators to obtain an estimate of the global or rigid body rotational motion,
\begin{eqnarray}
{\dot \theta}_{\rm est}(t) &=& \sum_{n =1}^N \dot\theta_{\rm est}^{(n)} (t),
\label{correlator-manycells}
\end{eqnarray}
where $N$ is the total number of photoreceptors and the local estimators  $\dot\theta^{(n)}$ at each point $n$ along the retina are defined in Eq. (\ref{fourpieces}); we have written this as if all the photoreceptors are arrayed along a line, but there is a simple generalization to a fully two dimensional array.
This computation takes as input not $2\times 50$ samples of the movie that we project onto the retina, but rather $N\times 50$, which can reach $D\sim 10^4$.  In this case the essential reduction of dimensionality is from this enormous space to one of only $(N/2)\times 4 \sim 10^2$ dimensions.  While dimensionality reduction still is a key to understanding the computation in this case, we would like to start with a more modest problem.
In principle we can probe the response of this estimator by stimulating only two photoreceptors \cite{franceschini+al_89}.
Alternatively we can limit the dimensionality of the input by restricting stimuli to a single spatial frequency, or equivalently just two components which vary as sine and cosine of the visual angle $x$,
\begin{equation}
I(x,t) = {\bar I}\cdot [1 + s(t) \sin(kx) + c(t)\cos(kx)] ,
\label{stimdef'n}
\end{equation}
where $I(x,t)$ is the light intensity with mean ${\bar I}$, $k/2\pi$ is the spatial frequency,
and the dynamics of the stimulus is defined by the functions $s(t)$ and $c(t)$.  The prediction of the correlator model Eq. (\ref{correlator-manycells}) is that the motion estimate is again determined by only four stimulus parameters, and in the limit that the cell integrates over a large number of receptors we find the simple result
\begin{eqnarray}
{\dot\theta}_{\rm est}(t) &\propto &
\left[ \int d\tau f(\tau) s(t-\tau) \right] 
\nonumber\\
&&\,\,\,\,\,\times \left[ \int d\tau' g(\tau ') c(t-\tau ') \right]
\nonumber\\
&&\,\,\,\,\,\,\,\,\,\,
- \left[ \int d\tau f(\tau) c(t-\tau)  \right] \nonumber\\
&&\,\,\,\,\,\,\,\,\,\,\,\,\,\,\,\times \left[ \int d\tau' g(\tau ') c(t-\tau ') \right] .
\end{eqnarray}
We emphasize that even with this simplification of the stimulus the known combination of temporal precision and integration times in motion computation mean that $\sim 10^2$ samples of the functions $s(t)$ and $c(t)$ could be relevant to the probability of spike generation in the motion sensitive neurons.  

Thus far we have discussed models of how the visual system could estimate motion; one can also ask if there is a way that the system {\em should} estimate motion.  In particular, for the problem of wide field motion estimation faced by the fly, we can ask how to process the signals coming from the array of photodetectors so as to generate an estimate of velocity which is as accurate as possible given the constraints of noise and blur in the photoreceptor array.  This is a well posed theoretical problem, and the results are as follows  \cite{potters+bialek_94}:

{\em Low SNR.} In the limit of low signal--to--noise ratios (SNR), the optimal estimator is a generalization of the correlator model in Eq. (\ref{correlator-manycells}),
\begin{eqnarray}
\dot\theta_{\rm est} (t) &=& \sum_{\rm nm} \int d\tau d\tau' V_{\rm n} (t-\tau ) K_{\rm nm} (\tau, \tau' ) V_{\rm m} (t-\tau ') \nonumber\\
&&\,\,\,\,\,\,\,\,\, + \cdots ,
\label{corr-opt}
\end{eqnarray}
where the higher order terms $\cdots$ include more complicated products of receptor voltages.  More precisely, this is the leading term in a power series expansion, and at low SNR the leading term is guaranteed to dominate.
The detailed structure of the kernels $K_{\rm nm}$ depend on our assumptions about the statistical structure of the visual world, but the general correlator form  is independent of these assumptions.  

{\em Intermediate SNR.} As the signal--to--noise ratio increases, higher order terms in the expansion of Eq (\ref{corr-opt}) become important and also the kernels $K_{\rm nm}$ become modified so that the optimal estimator integrates over shorter times when the signals are more robust and when typical velocities are larger.

{\em High SNR.}  At high SNR, under a broad range of assumptions about the statistics of the visual world the optimal estimator crosses over to approximate
\begin{eqnarray}
\dot\theta_{\rm est} (t) &=& {{\sum_{\rm n} [V_{\rm n+1}(t) - V_{\rm n}(t)] \cdot (dV_{\rm n}(t)/dt)}
\over{A + \sum_{\rm n}  [V_{\rm n+1}(t) - V_{\rm n}(t)] ^2}}
\label{opt-highsnr1}\\
&\approx& {{\int dx [\partial C(x,t)/\partial x]\cdot [\partial C(x,t)/\partial t]}
\over{A' + \int dx [\partial C(x,t)/\partial x]^2}}
\label{opt-highsnr2}\\
&\rightarrow& {{\partial C(x,t)/\partial t}\over{\partial C(x,t)/\partial x}}
\label{opt-highsnr3}
\end{eqnarray}
where $A$ and $A'$ are constants that depend on the details of the image statistics and the last expression indicates schematically the limiting behavior at high contrast.  In this limit the optimal estimator is just the ratio of temporal and spatial derivatives.  At very high SNR there is no need to average over time to suppress noise, so we show the estimator as being an instantaneous function of the receptor voltages and their derivatives; more generally at finite SNR the form of the estimator is the same but the receptor responses need to be smoothed over time.

Perhaps the most interesting result of the theory is that both the correlator model and a ratio of derivatives model emerge as opposite limiting cases of the general estimation problem.  The ratio of derivatives is in some sense the naive solution to the problem of motion estimation, since if the movie on the retina has the form $C(x,t) = F(x-vt)$, corresponding to pure rigid motion, then it is easy to see that the velocity $v = - (\partial C/\partial t) /(\partial C/\partial x)$.  This isn't the general solution because the combination of  differentiation and division tends to amplify noise; thus the ratio of derivatives emerges only as the high SNR limit of the general problem.  At the opposite extreme, the correlator model is maximally robust against noise although it does make well known systematic errors by confounding the true velocity with the contrast and spatial structure of the image; the theory shows that these errors---which have correlates in the responses of the motion sensitive neurons, as well as in behavioral experiments---may emerge not as limitations of neural hardware but rather as part of the optimal strategy for dealing with noise.

Although the theory of optimal estimation gives a new perspective on the correlator model, one can hardly count the well established correlator--like behavior of motion sensitive neurons as a success of the theory.  The real test would be to see the crossover from correlator--like behavior to the ratio of derivatives.  Notice from Eq. (\ref{opt-highsnr2}) that if the overall SNR is large but the contrast is (instantaneously) small, the optimal estimate is again correlator--like because the contrast dependent term in the denominator can be neglected.  Thus even under a statistically stationary set of conditions corresponding to high SNR, we should be able to see both the correlator, with its mutiplicative nonlinearity, and the divisive nonlinearity from the ratio of derivatives.  Behaviors consistent with this prediction have been observed \cite{ruyter+al_94},  but we would like a more direct demonstration.  

If we consider stimuli of the simple form in Eq. (\ref{stimdef'n}), then it is easy to see that the motion estimator in Eq. (\ref{opt-highsnr2}) can be written as
\begin{equation}
\dot\theta_{\rm est}(t) \approx {{s(t)\cdot [dc(t)/dt] - c(t)\cdot [ds(t)/dt]}\over
{B + [s^2(t) + c^2(t)]}},
\end{equation}
where again $B$ is a constant.  More generally, if the receptor signals are all smoothed in the time domain by a filter $f(\tau )$, then by analogy with Eq's. (\ref{s_1}--\ref{s_4}), we can define four relevant dimensions of the input movie,
\begin{eqnarray}
s_1 &=& \int d\tau f(\tau) s(t-\tau )
\label{s1_final}\\
s_2 &=& \int d\tau f(\tau) c(t-\tau )\\
s_3 &=& \int d\tau f(\tau) {{ds(t-\tau )}\over{dt}} \nonumber\\
&=& \int d\tau {{df(\tau)}\over{d\tau}} s(t-\tau)\\
s_4 &=& \int d\tau f(\tau) {{dc(t-\tau )}\over{dt}} \nonumber\\
&=& \int d\tau {{df(\tau)}\over{d\tau}} c(t-\tau),
\label{s4_final}
\end{eqnarray}
and then 
\begin{equation}
\dot\theta_{\rm est}(t) \approx {{s_1 \cdot s_4- s_2 \cdot s_3}\over
{B + [s_1^2 + s_2^2]}}.
\label{optest-final}
\end{equation}
Thus the optimal estimator again is a function of four relevant dimensions out of the high dimensional space of input signals, these dimensions are built by operating on $s(t)$ and $c(t)$ with two filters where one is the time derivative of the other, and then the four dimensions are combined nonlinearly.  By analogy with Eq. (\ref{fourpieces}), we can identify the ``correlator variable''  ${\cal V}_{\rm corr}$ constructed from  these four variables,
\begin{equation}
{\cal V}_{\rm corr} = s_1 \cdot s_4 - s_2 \cdot s_3 ,
\label{Vcorr}
\end{equation}
and the full optimal estimator normalizes this correlator variable through a divisive nonlinearity,
\begin{equation}
{\dot\theta }_{\rm est} = { {{\cal V}_{\rm corr} }\over{B + {\cal D }  } },
\label{definesD}
\end{equation}
where ${\cal D} = s_1^2 + s_2^2$  approximates the mean  square spatial derivative of the image. Note that the different dimensions enter in highly symmetric combinations.

The goal of the experiments described below is in fact to test the predictions from Eq. (\ref{optest-final}), but we can also view this as an example of the more general problem of searching for relevant low dimensional structures within the high dimensional space of inputs to which a neuron responds.  Related combinations of multiplicative and divisive nonlinearities arise quite generally in models for the ``normalization'' of neural responses in visual cortex \cite{heeger_92,carandini+al_97}  and in particular in the normalization models applied to the motion sensitive cells of primate area MT \cite{simoncelli+heeger_98}. Recent work \cite{weiss+al_02} suggests that this sort of computation can be derived for motion estimation in primate vision from the same sorts of optimization arguments used previously for insect vision \cite{potters+bialek_94}.

Although our focus here has been on the problem of motion estimation in vision, it is important to note that the same kind of dimensionality reduction provides a precise formulation of feature selectivity in other systems as well.  The classic example of center--surround organization in retinal ganglion cells \cite{barlow_53} can be thought of as projecting the image (or movie) onto two dimensions corresponding to spatial filters with different radii.  Truly linear center--surround behavior would then correspond to the cell responding to just a linear combination (difference) of these two dimensions, so that really only one combined projection is relevant, while more subtle forms of interaction between center and surround (e.g., shunting inhibition) would still correspond to a projection onto two dimensions but the nonlinear operation with relates firing probability to location in this two dimensional space would be more complicated.  Similarly, the oriented receptive fields in cortical simple cells are described as having multiple subregions \cite{hubel+wiesel_62}, but if there are nonlinear interactions among the subregions then effectively there is no single projection of the stimulus to which the cell responds; rather the small number of subregions defines a projection onto a low dimensional subspace of images and there may be nontrivial computations within this subspace.  Indeed, computational models for the detection of contours and object boundaries have precisely this sort of nonlinear interaction among linear receptive subfields \cite{iverson+zucker_95}.  

More generally, while filtering and feature selectivity sometimes are used as synonyms, actually {\em detecting} a feature requires a logical operation, and in fact interesting features may be defined by conjunctions of logical operations.  These more sophisticated notions of feature selectivity are not summarized  by a single filter or receptive field.   These computations do fit, however, within the framework suggested here, as nonlinear (perhaps even hard threshold) operations on a low dimensional projection of the stimulus.

\section{Searching for low dimensional structures}
\label{methods}

We have argued that models of motion estimation, and perhaps other examples of neural fetaure selectivity, belong to a class of models  in which neurons are sensitive only to a low dimensional
projection of their high dimensional input signals.  One approach would
be to find the best model in this class to describe particular neurons.  
It would, however, be more compelling if we could provide direct evidence for
applicability of the {\em class} of models before finding the best model
within the class.  
The essential hypothesis of Eq. (\ref{modeldef}) is that neurons are
sensitive to a subspace of inputs with dimensionality $K$ much smaller than
the full stimulus dimensionality $D$.  This suggests   a series of questions:
\begin{enumerate}
\item Can we make a direct measurement of $K$, the number of relevant
dimensions?
\item Can we find the set of vectors $\{{\bf\hat v}_{\rm n}\}$ that span
this relevant subspace?
\item Can we map the nonlinearity $G(\cdot)$ that the neuron implements
within this space?
\end{enumerate}
In the simple case where there is only one relevant dimension, the idea of reverse or
triggered correlation \cite{boer+kuyper_68, rieke+al_97} allows us to find this one
special direction in stimulus space provided that we choose our ensemble of stimuli
correctly.  If we want to test a model in
which there are multiple stimulus dimensions we need to compute objects that have a
chance of defining more than one relevant vector.   The basic suggestion comes
from early work on the  fly's motion sensitive visual neurons \cite{ruyter+bialek_88}.
Instead of  computing the average stimulus that precedes a spike, we can characterize the fluctuations around
the average by their covariance matrix.  Along most directions in stimulus space, this covariance matrix has a
structure determined only by correlations in the stimulus.  There are a small number of directions, however, along which the stimuli that trigger spikes have a different variance than expected a priori.  The fact that the number of directions with different variances is small provides direct evidence that the cell is sensitive only to a small number of projections. Further, identifying the directions along which the variance is different provides us with a coordinate system that spans the set of relevant projections.
The following arguments, leading to Eq. (\ref{deltaCresult}),   formalize this
intuition, answering the first two questions above.  
Then we turn to an analysis of the nonlinearities in $G$, leading to Eq. (\ref{rfromsamples}).

It is useful to think about the spike train  as a sum of
unit impulses,
\begin{equation}
\rho(t) = \sum_{\rm i} \delta(t - t_{\rm i}) ,
\end{equation}
where the $t_{\rm i}$ are the spike times.  Then the quantities
of interest are correlation functions between $\rho (t)$ and the stimulus vector
${\bf\vec s}_t$;  recall that this vector can represent both the spatial and temporal
variations in the stimulus movie. As an example, the average stimulus preceding a
spike is
\begin{equation}
\langle {\bf\vec s}_{t_{\rm spike}}\rangle
= {1\over {\bar r}}
\langle \rho(t) {\bf\vec s}_t\rangle  ,
\label{sta}
\end{equation}
where $\bar r$ is the mean spike rate and $\langle \cdots \rangle$ denotes an
average over a very long experiment.  If we repeat the same stimulus for many trials
and average the resulting spike trains, then we will obtain the probability per unit
time $r(t)$ that the cell spikes, where $t$ is measured by a clock synchronized with
the repeating stimulus as in the usual poststimulus time histogram.
Thus
\begin{equation}
\langle \rho(t) \rangle_{\rm trials} = r(t) .
\end{equation}
The spike rate $r(t)$ is an average of the spike
train $\rho(t)$ over all the noise in the neural response, so that
when we need to compute averages over a long experiment, we imagine doing this
formally by first  averaging over the noise with the stimulus held fixed, and then
averaging over the distribution of signals; for example,
\begin{equation}
\langle \rho(t) {\bf\vec s}_t\rangle =  \langle r(t) {\bf\vec s}_t\rangle_{s}  ,
\end{equation}
where $\langle \cdots \rangle_s$ denotes an
average over the distribution of signals presented in the
experiment.

To find multiple relevant directions we consider the matrix of second moments that characterizes the stimuli leading to a spike \cite{ruyter+bialek_88}.  If the components of the stimulus vector ${\bf \vec s}_t$ are written as ${\bf s}_t(i)$, with the index $i=1,2, \cdots , D$ running over the full dimensionality of the stimulus space, then the second moments of stimuli preceeding
a spike are
\begin{eqnarray}
C_{\rm spike}(i,j) &\equiv& \langle {\bf s}_{t_{\rm spike}} (i) 
{\bf s}_{t_{\rm spike}} (j)\rangle
\label{C-empirical}
\\
&=& {1\over {\bar r}} \langle
\rho(t) {\bf s}_t(i) {\bf s}_t(j)\rangle .
\end{eqnarray}
From the arguments above this can be rewritten as
\begin{equation}
C_{\rm spike}(i,j) = {1 \over {\bar r}} \langle r(t) {\bf s}_t (i) 
{\bf s}_t (j)\rangle .
\label{C-rates}
\end{equation}
It is crucial that $C_{\rm spike}(i,j)$ is something we can estimate directly from data,
looking back at the stimuli that lead to a spike and computing the matrix of their
second moments according to the definition in Eq. (\ref{C-empirical}).  On the other
hand, Eq. (\ref{C-rates}) gives us a way of relating these computations from the data
to underlying models of how the spike rate $r(t)$ depends on the stimulus.

In general it is hard to go further than Eq. (\ref{C-rates}) analytically.  More precisely,
with stimuli chosen from an arbitrary distribution the relation between $C_{\rm
spike}$ and some underlying model of the 
response can be arbitrarily complicated \cite{sharpee+al_04}.  We can make
progress, however, if we are willing to restrict our attention to stimuli that are drawn
from a Gaussian distribution as in reverse correlation analyses.  It is important to realize that this restriction, while significant, does not specify a uniquely ``random'' stimulus.  Gaussian does not imply white; we can construct an arbitrary correlation function for our
stimuli, including correlation functions modelled after natural signals \cite{rieke+al_95}.  Further, we can construct stimuli which are nonlinear
functions of underlying ``hidden'' Gaussian variables; these stimuli can have a complex
and even naturalistic structure---see, for example, Ref \cite{fairhall+al_01}---and such hidden
variable methods may be useful as a bridge to more general application of the dimensionality
reduction idea.

If the distribution of signals is Gaussian, then
averages such as Eq. (\ref{C-rates}) are straightforward to compute.
The key step is the following identity:  If ${\bf \vec x} = x_1, x_2, \cdots , x_D$ is a vector drawn from a multidimensional Gaussian distribution with zero mean, and $f({\bf \vec x})$ is a differentiable function of this vector, then
\begin{equation}
\langle x_{\rm i} f({\bf \vec x}) \rangle = \sum_{{\rm j}=1}^D C_{\rm ij} {\Bigg\langle} {{\partial f({\bf \vec x})}\over{\partial x_{\rm j}}} {\Bigg\rangle} ,
\end{equation}
where $C_{\rm ij} = \langle x_{\rm i} x_{\rm j}\rangle$ is the covariance matrix of $\bf\vec x$.  This can be applied twice:
\begin{eqnarray}
\langle x_{\rm i} x_{\rm  j} f({\bf \vec x}) \rangle &=& 
\sum_{{\rm k}=1}^D C_{\rm ik} 
{\Bigg\langle} {{\partial [x_{\rm j}f({\bf \vec x})]}\over{\partial x_{\rm k}}} {\Bigg\rangle} \nonumber\\
&=&  
\sum_{{\rm m}=1}^D C_{\rm im} \left[\delta_{\rm jm} \langle f({\bf \vec x}) \rangle
+ 
{\Bigg\langle} x_{\rm j} {{\partial f({\bf \vec x})}\over{\partial x_{\rm m}}}
{\Bigg\rangle}\right]
\nonumber\\
&&
\\
&=&
C_{\rm ij} \langle f({\bf \vec x}) \rangle
+ \sum_{{\rm n,\,m}=1}^D 
C_{\rm im} C_{\rm jn}
{\Bigg\langle}
{{\partial^2 f({\bf \vec x})}\over{\partial x_{\rm m} \partial x_{\rm n}}}
{\Bigg\rangle} .\nonumber\\
&&
\end{eqnarray}
We can use this in evaluating $C_{\rm spike}$ from Eq (\ref{C-rates}) by identifying the vector $\bf\vec x$ with the stimulus ${\bf \vec s}_t$ and the spike rate $r(t)$ with the function $f({\bf \vec x})$.  The result is
\begin{eqnarray}
C_{\rm spike}(i,j) &=& C_{\rm prior} (i,j) + \Delta C (i,j),\\
\Delta C (i,j) &=&
{1\over {\bar r}} 
C_{\rm prior} (i,k)
\Bigg\langle
{{\partial^2 r(t)}\over{\partial {\bf s}_t(k) \partial {\bf s}_t(l)}}
\Bigg\rangle
C_{\rm prior} (l,j),\nonumber\\
&&
\end{eqnarray}
where we sum over the repeated indices $k$ and $l$, and $C_{\rm prior}(i,j)$ is the second moment of stimuli averaged over the whole
experiment,
\begin{equation}
C_{\rm prior} (i,j) = \langle {\bf s}_t(i) {\bf s}_t(j)\rangle .
\end{equation}
Further, if the rate has the `low dimensional' form
of Eq. (\ref{modeldef}), then the derivatives in the full stimulus space reduce to
derivatives of the function $G$ with respect to its $K$ arguments:
\begin{eqnarray}
{{\partial^2 r(t)}\over{\partial{\bf s}_t (k)\partial{\bf s}_t (l)}}
&=&
{\bar r}
{{\partial^2 G(s_1, s_2, \cdots s_K)}\over{\partial s_\alpha \partial s_\beta}} 
{\bf v}_\alpha(k) {\bf v}_\beta(l) ,\nonumber\\
&&
\end{eqnarray}
where as with the stimulus vector ${\bf \vec s}_t$ we use ${\bf v}_\alpha(i) $
to denote the components of the projection vectors ${\bf \vec v}_\alpha$; again the
index
$i$ runs over the full dimensionality of the stimulus, $i=1,2, \cdots, D$ while the index
$\alpha$ runs over the number of {\em relevant} dimensions, $\alpha = 1, 2, \cdots , K$, and we sum over repeated indices $\alpha$ and $\beta$. 

Putting these results together,  we find an expression for the difference
$\Delta C$ between the second moments of stimuli that lead to a spike and stimuli
chosen at random:
\begin{eqnarray}
\Delta C (i,j) &=&
\left[
C_{\rm prior} (i,k){\bf v}_\alpha(k)\right]
A (\alpha,\beta)\nonumber\\
&&\,\,\,\,\,\times
\left[
{\bf v}_\beta(l) C_{\rm prior} (l,j)\right] ,
\label{deltaCresult}\\
A (\alpha,\beta) &=&
\Bigg\langle
{{\partial^2 G(s_1, s_2, \cdots s_K)}\over{\partial s_\alpha
\partial s_\beta}} 
\Bigg\rangle ,
\end{eqnarray}
and we sum over all repeated indices $\alpha,\,\beta,\,k$ and $l$ in Eq. (\ref{deltaCresult}).
There are several important points which follow from these expressions.

First, Eq. (\ref{deltaCresult}) shows that $\Delta C(i,j)$, which is a $D\times D$ matrix,
is determined by  the $K\times K$ matrix $A(\alpha,\beta)$ formed from the second derivatives of the function $G$. This means that
$\Delta C(i,j)$ can have only $K$ nonzero eigenvalues, where $K$ is the number of relevant stimulus dimensions.  Thus we can test directly the hypothesis that the number of relevant dimensions is
small just by looking at the eigenvalues of  $\Delta C$.  Further, this test is
independent of assumptions about the nature of the nonlinearities represented by
the function $G$.

Second,  the eigenvectors of $\Delta C$ associated with the nonzero eigenvalues are linear
combinations of the vectors ${\bf \vec v}_\alpha$, blurred by the correlations in the
stimulus itself.  More precisely, if we look at the set of nontrivial eigenvectors ${\bf
\vec u}_{\alpha}$, with $\alpha = 1, 2, \cdots ,K$, and undo the effects of
stimulus correlations to form the vectors 
${\bf \vec v}_\alpha' = [C_{\rm prior}]^{-1} {\bf \cdot} {\bf \vec u}_{\alpha}$, 
then we will find that these vectors span the same space as
the vectors ${\bf\vec v}_{\alpha}$ which define the relevant subspace of stimuli.

Third, we note that the  eigenvalue analysis of  $\Delta C$ is {\em not} a principal components analysis of the stimulus probability distribution.  In particular, unless the function $G$ were of a very special form, the distribution of stimuli that lead to a spike will be strongly non--Gaussian, and so a principal components analysis of this distribution will not capture its structure.  Further, directions in stimulus space that have small variance can nonetheless make large contributions to $\Delta C$.  Note also that the eigenvalues of $\Delta C$ can be both positive or negative, while of course the spectrum of a covariance matrix (associated with the principal components of the underlying distribution)  always is positive.

Finally,  the eigenvectors (or their deblurred versions) that emerge from this analysis are  useful  {\em only}  because they define a set of dimensions spanning the space of relevant stimulus features.  Once we are in this
restricted space, we are free to choose any set of coordinates.
In this sense, the notion of finding ``the'' linear filters or
receptive fields that characterize the cell becomes meaningless
once we leave behind a model in which only one stimulus dimension
is relevant.   The only truly
complete characterization is in terms of the full nonlinear
input/output relation within the relevant subspace.

Once we have identified a subspace of stimuli $s_1 , s_2, \cdots , s_K$, we actually can map the nonlinear function $G$ directly provided that $K$ is not too large.  We recall that the spike rate $r(t)$ is the probability per unit time that a spike will occur at time $t$, given the stimulus ${\bf \vec s}_t$ leading up to that time.  Formally,
\begin{equation}
r(t) = P[{\rm spike} \,@\,t|{\bf \vec s}_t]  .
\end{equation}
From Eq. (\ref{modeldef}), the rate depends only on $K$ projections of the stimulus, and so
\begin{equation}
r(t) = P[{\rm spike} \,@\,t| s_1 , s_2, \cdots , s_K ]  .
\end{equation}
But the probability of a spike given the stimulus can be rewritten using Bayes' rule:
\begin{eqnarray}
P[{\rm spike} \,@\,t  {\hskip-0.12in}&|& {\hskip-0.12in} s_1 , s_2, \cdots , s_K ] = {{P[{\rm spike} \,@\,t ] }\over{P[ s_1 , s_2, \cdots , s_K]}}
 \nonumber\\
&\times&  P[ s_1 , s_2, \cdots , s_K | {\rm spike} \,@\,t] .
\end{eqnarray}
In the same way that the function $P[{\rm spike} \,@\,t|{\bf \vec s}_t] $ gives the time dependent spike rate $r(t)$, the number $P[{\rm spike} \,@\,t]$ is just the average spike rate $\bar r$.  Thus the nonlinear computation within the $K$--dimensional relevant subspace that determines the neural response can be found from the ratio of probability distributions in this subspace, 
\begin{eqnarray}
r(t)  &=& {\bar r}  G( s_1 , s_2, \cdots , s_K )\\
& =& {\bar r}  \cdot {{P[ s_1 , s_2, \cdots , s_K | {\rm spike} \,@\,t] }\over{P[ s_1 , s_2, \cdots , s_K ]}}.
\label{rfromsamples}
\end{eqnarray}
Now the full distribution $P[ s_1 , s_2, \cdots , s_K ]$ is known, since this defines the conditions of the experiment; further, we have considered situations in which this distribution is Gaussian and hence is defined completely by a $K\times K$ covariance matrix.  The probabiliity distribution of stimuli given a spike, the response--conditional ensemble \cite{ruyter+bialek_88}, can be estimated by sampling:  each time we see a spike, we can look back at the full stimulus ${\bf\vec s}_t$ and form the $K$ projections $s_1, s_2, \cdots , s_K$; this set of projections at one spike time provides one sample drawn from the distribution $P[ s_1 , s_2, \cdots , s_K | {\rm spike} \,@\,t] $, and from many such samples we can estimate the underlying distribution.  This Bayesian strategy for mapping the nonlinear input/ouput relation provides a large dynamic range proportional to the total number of spikes observed in the experiment---see, for example, Ref. \cite{brenner+al_00}.

We emphasize that the procedure described above rests not on a family of parameterized models which we fit to the data, but rather on the idea that if the dimensionality of the relevant subspace is sufficiently small then we don't really need a model.  In practice we have to assume only that the relevant functions are reasonably smooth, and then the combination of eigenvalue analysis and Bayesian sampling provides explicit answers to the three questions raised at the beginning of this section.

\section{An experiment in H1}
\label{h1expt}

\begin{figure}[t]
\centering
\epsfxsize=\linewidth
\epsfbox{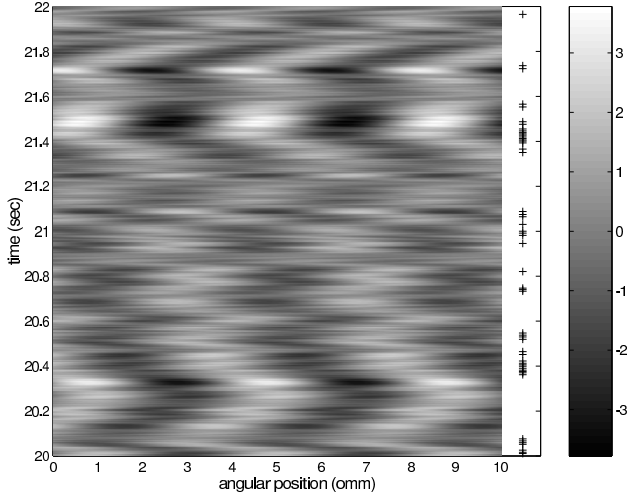}
\caption{A two second segment of the stimulus and corresponding neural response.
The stimulus movie consists of vertical stripes.  Here each horizontal slice
indicates the pattern of these stripes at one instant of time.  Plus signs at
the right indicate spike times from H1.  Brief periods with  coherent
motion to the left are correlated with the spikes, while clear motions to the
right inhibit the neuron.  The challenge of the present analysis is to make
more precise this connection between features of the movie and the probability
of spiking.}
\label{stimandspikes}
\end{figure}

We apply these ideas to an experiment on the fly's motion sensitive H1 neuron, where
we would like to dissect the different features of motion computation discussed in
Section \ref{models}.  A segment of the visual stimulus and H1's response is shown in Fig. \ref{stimandspikes}.  While the most elementary model of motion computations involves temporal
comparisons between two pixels, H1 is a wide field neuron and so is best stimulated
by spatially extended stimuli.  To retain the simplicity of the two--pixel limit in an
extended stimulus we consider here a stimulus which has just one spatial frequency, as in Eq. (\ref{stimdef'n}).  For technical details of stimulus generation and neural recording see Appendix \ref{app-expt}.

\begin{figure}[b]
\centering
\epsfxsize=\linewidth
\epsfbox{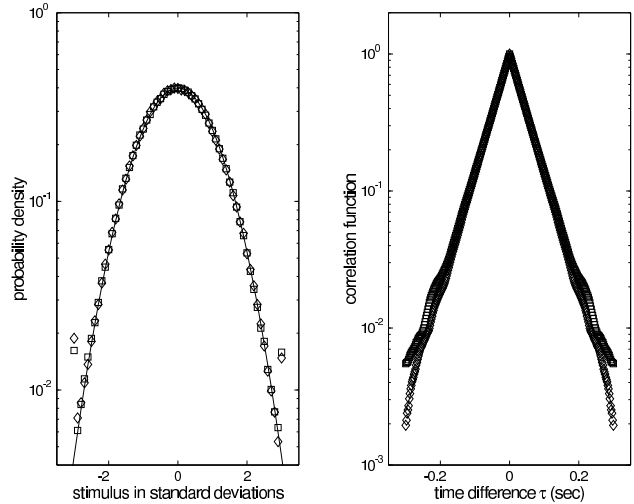}
\caption{Statistics of the visual stimulus.  Probability distribution of $s(t)$ and $c(t)$ compared with a Gaussian.  Because these signals represent image contrast, there is inevitably some clipping at large amplitude (light intensity cannot be negative), but this affects only $\sim 1\%$ of the signals.  At right, the autocorrelation of the signals, $\langle s(t) s(t+\tau)\rangle$ (circles) and $\langle c(t) c(t+\tau)\rangle$ (squares) are almost perfect exponential decays $\exp(-|\tau |/\tau_c)$,  $\tau_c = 50\,{\rm msec}$.}
\label{stimstats}
\end{figure}

To make use of the results derived above, we choose $s(t)$ and $c(t)$ to be Gaussian
stochastic processes, as seen in Fig, \ref{stimstats}, with correlation times of 50
msec.  Similar experiments with correlation times from 10--100 msec lead to
essentially the same results described here, although there is 
adaptation to the correlation time, as expected from earlier work \cite{ruyter+al_96,ruyter+al_94}.
The problem we
would like to solve is to describe the relation between the stimulus movie $I(x,t)$ and
the spike arrival times (cf. Fig. \ref{stimandspikes}).

Simple computation of the spike triggered average movie produces no statistically
significant results:  the cell is sensitive to motion, and invariant to the overall
contrast of the movie, so that the stimulus generated by the transformation $(s,c)
\rightarrow (-s,-c)$ will produce indistinguishable responses.  This being said, we
want to proceed with the covariance matrix analysis.

\begin{figure}[t]
\centering
\epsfxsize=\linewidth
\epsfbox{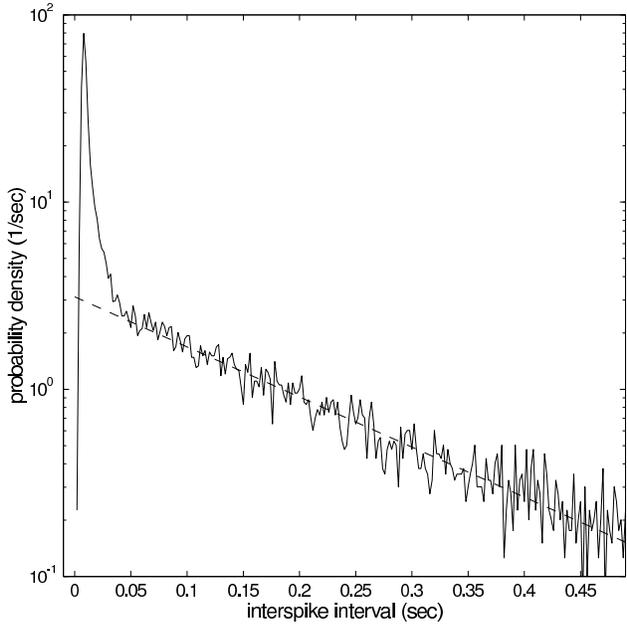}
\caption{Distribution of interspike intervals.  Solid line shows the observed
distribution collected in 2 msec bins.  Dashed line is an exponential fit to
the tail of the distribution.  Exponential decay of the interval distribution
means that successive spikes are occurring independently, and the data indicate
that one must wait $\sim 40\,{\rm msec}$ for this independence to be
established.  Further analysis is done only with these ``isolated spikes.''}
\label{isis}
\end{figure}

From previous work  we know that different patterns of spikes in time can stand for very different motion trajectories \cite{ruyter+bialek_88}.    From the present point  of view, this connection to spike patterns is not a question about the nature of the motion computation but rather about how the output of this computation is represented in the spike train.  To simplify the discussion, we will focus on spikes that occur in relative isolation from previous spikes.  Specifically, when we look at the interspike interval distribution in Fig. \ref{isis}, we see that for intervals longer than $\sim 40\,{\rm msec}$ the distribution has the form of a decaying exponential.  This is what we expect if after such long intervals spikes are generated independently without memory or correlation to previous spikes.  More colloquially, spikes in this exponential regime are being generated independently in response to the stimulus, and not in relation to previous spikes.  
All further analysis is done using these isolated spikes; for a related discussion in the context of model neurons see Ref. \cite{aguera+al_03}.

\begin{figure}[b]
\centering
\epsfxsize=\linewidth
\epsfbox{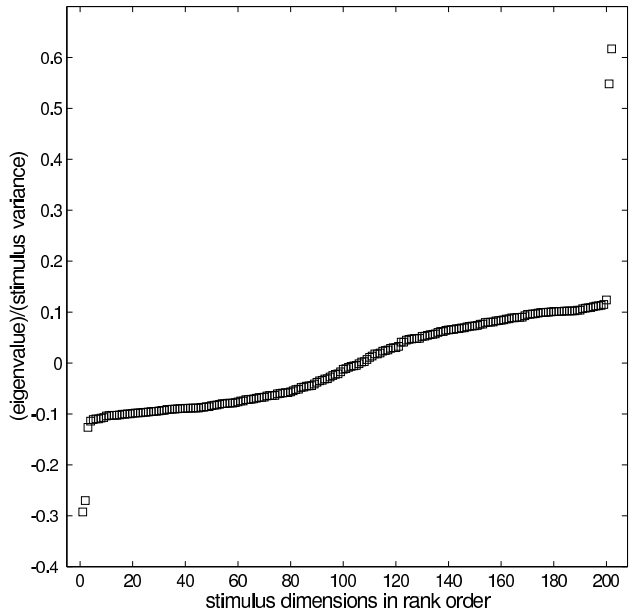}
\caption{Eigenvalues of $\Delta C$.  Stimuli are represented as segments of $s(t)$ and $c(t)$ in windows of $\pm 200\,{\rm msec}$ surrounding isolated spikes, sampled at $4\,{\rm msec}$ resolution; $\Delta C$ thus is a $200\times200$matrix.  As explained in
the text, the eigenvalue measures the spike--triggered change in stimulus
variance along a particular direction in stimulus space, while the eigenvector
specifies this direction.  Since the stimulus itself has correlations,
different directions have different variances a priori, and we express the
change in variance as a fraction of the total variance along the corresponding
direction.  There are four dimensions which stand out clearly from the
background.}
\label{eigrats}
\end{figure}

Qualitative examination of the change in stimulus covariance in the neighborhhod of an isolated spike, $\Delta C$, reveals that it is a very smooth matrix, consistent with the idea that it is composed out of a small number of significant eigenvectors.
To quantify these observations, and proceed along the analysis program outlined in the previous section, we diagonalize $\Delta C$ to give the eigenvalues shown in 
%
Fig. \ref{eigrats}.   In trying to plot the results there is a natural question about units.  Because the stimuli themselves are not white, different stimulus components have different variances.  The eigenvalue analysis of $\Delta C$ provides a  coordinate system on the stimulus space, and the eigenvalues themselves measure the change in stimulus variance along each coordinate when we trigger on a spike.  Small changes in variance along directions with large total variance presumably are not very meaningful, while along a direction with small variance even a small change could mean that the spike points precisely to a particular value of that stimulus component.  This suggests measuring the eigenvalues of $\Delta C$ in units of the stimulus variance along each eigendirection, and this is what we do in Fig. \ref{eigrats}.  This normalization has the added value (not relevant here) that one can describe the stimulus in terms of components with different physical units and still make meaningful comparisons among the different eigenvalues.
Figure \ref{eigrats} shows clearly that four directions in stimulus space stand out  relative to a background of 196 other dimensions.  The discussion in Section \ref{models} of models for motion estimation certainly prepares us to think about four special directions, but before looking at their structure we should answer questions about their statistical significance.  

\begin{figure}[t]
\centering
\epsfxsize=\linewidth
\epsfbox{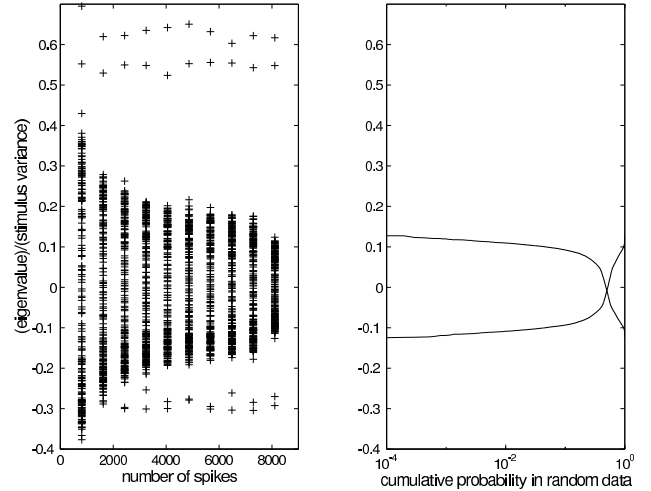}
\caption{Testing the significance of the eigenvalue distributions.
At left we show the evolution of the eigenvalue spectrum as we analyze larger
data sets.  The four dimensions which stand out from the background in the
full data set also have stable eigenvalues as a function of data set size, in
contrast to the background of eigenvalues which come from a distribution which
narrows as more data is included.  At right we show the cumulative probability
distribution of eigenvalues from surrogate data in which the correlatons between
stimulus and spike train have been broken by random time shifts, as explained
in the text. Eigenvalues with absolute value larger than $0.1$
arise roughly $1\%$ of the time, but there is a very steep edge to the
distribution such that absolute values larger than $0.13$ occur only $0.01\%$
of the time.  The sharp edge in the random data is essential in identifying
eigenvalues which stand out from the background, and this edge is inherited
from the simpler problem of eigenvalues in truly random matrices \cite{mehta_67}.}
\label{significance}
\end{figure}

In practice we form the matrix $\Delta C$ from a finite amount of data; even if spikes and stimuli were completely uncorrelated, this finite sampling  gives rise to some structure in $\Delta C$ and to a spectrum of eigenvalues which broadens around zero.  One way to check the significance of eigenvalues is to examine the dependence of the whole spectrum on the number of samples.  Out of the $\sim 8000$ isolated spikes which we have collected in this experiment, we show in left panel of Fig. \ref{significance} what happens if we choose 10\%, 20\%, ... , 90\% at random to use as the basis for constructing $\Delta C$.  The basic structure of four modes separated from the background is clear once we have included roughly half the data, and the background seems (generally)  to narrow as we include more data.  

A different approach to statistical significance is to generate a set of random data that have comparable statistical properties to the real data, breaking only the correlations between stimuli and spikes.  If we shift the all the spikes forward in time by  several seconds relative to the stimulus, then since the correlations in the stimulus itself are short--lived, there will be no residual correlation between stimulus and spikes, but all the internal correlations of these signals are untouched.  If we choose the shift times at random, with a minimum value, then we can generate many examples of uncorrelated stimuli and spikes, and find the eigenvalue spectra of $\Delta C$ in each example.  Taken together a large number of these examples gives us the distribution of eigenvalues that we expect to arise from noise alone, and this distribution is shown in cumulative form in the right panel of Fig. \ref{significance}.  A crucial point---expected from the analytic analysis of eigenvalues in simpler cases of random matrices \cite{mehta_67}---is that the distribution of eigenvalues in the pure noise case has a sharp edge rather than a long tail, so  that the band of eigenvalues in a single data set will similarly have a fairly definite endpoint rather than a long tail of `stragglers' which could be confused with significant dimensions.  While larger data sets might reveal more significant dimensions, Fig. \ref{significance} indicates that the present data set points to four and only four significant stimulus dimensions out of a total of 200.

\begin{figure}[b]
\centering
\epsfxsize=\linewidth
\epsfbox{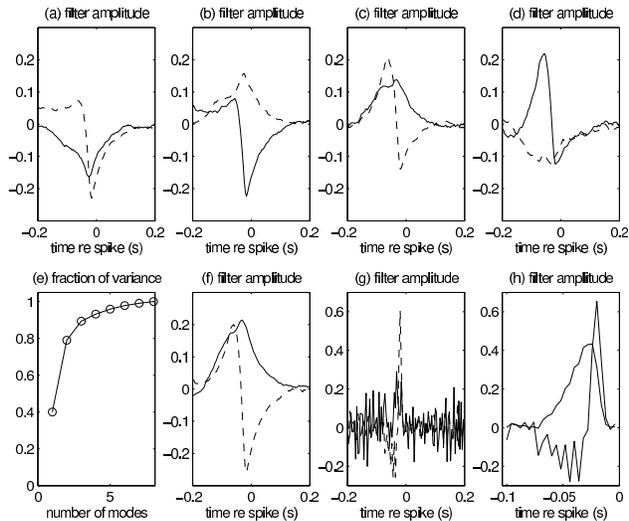}
\caption{Top panels: Eigenvectors associated with the four significant eigenvalues.
Solid lines show components along the $s$ stimulus directions, dashed lines
along the $c$ directions. Bottom panels: Analysis of filters.  (e) Results of singular value decomposition
demonstrate that most of the variation among the eight filters at left
can be captured by two modes, which are shown in (f). 
Deconvolving the stimulus correlations from these vectors we find the results
in (g), where we note that anti--causal pieces of the filters are now just
noise, as expected if the deconvolution is successful.  Closer examination
shows that one of these filters is approxmiately the derivative of the other,
and in (h) we impose this condition exactly and truncate the filters to the
$100\,{\rm msec}$ window which seems most relevant.}
\label{rawfilters}
\end{figure}

In Fig. \ref{rawfilters} we show the eigenvectors of $\Delta C$ associated with the four significant nonzero eigenvalues.  We can think of these eigenvectors as filters in the time domain which are applied to the two spatial components of the movie $s(t)$ and $c(t)$; the four eigenvectors thus determine eight filters.  We see that among these eight filters there are some similarities.  For example, the filter applied to $c(t)$ in eigenvector (a) looks very similar to that which is applied to $s(t)$ in eigenvector (b), the filter applied to $s(t)$ in eigenvector (c) is close to being the negative of that which is applied to $c(t)$ in eigenvector (d), and so on.  Some of these relations are a consequence of the approximate  invariance of H1's response to static translations of the visual inputs, and this is related to the fact that the significant eigenvalues form two nearly degenerate pairs.   In fact the similarity of the different filters is even greater than required by translation invariance, as indicated by the singular value decompostion shown in the lower panels of Fig. \ref{rawfilters}:  the eight temporal filters which emerge from the eigenvalue analysis are constructed largely from only two underlying filters which account for 80\% of the variance among the filter waveforms.    These two filters are extended in time because they still contain [as expected from Eq. (\ref{deltaCresult})] a ``blurring'' due to intrinsic correlations in the stimulus ensemble.  When we deconvolve these correlations we inevitably get a noisy result, but it is clear that the two filters form a pair, one which smooths the input signal over a $\sim 40$ msec window, and one which smooths the time derivative of the input signal over the same window.

The results thus far already provide confirmation for important predictions of the models  discussed in Section \ref{models}.   With stimuli of the form used here, these models predict that the motion estimator is constructed from four relevant stimulus dimensions,  that these dimensions in turn are built from  just two distinct temporal filters applied to two different spatial components of the visual input, and that one filter is the time derivative of the other [cf. Eq's. (\ref{s1_final}--\ref{optest-final})].  All three of these features are seen in Figs. (\ref{eigrats}) and (\ref{rawfilters}).

\begin{figure}[t]
\centering
\epsfxsize=\linewidth
\epsfbox{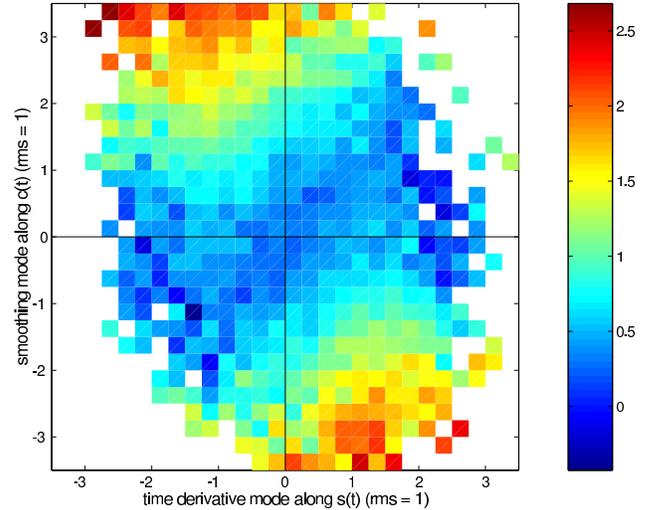}
\caption{Spike probability in two stimulus dimensions.  Color code represents
$\log_{10}[r(t)/({\rm spikes/sec)}]$, with $r(t)$ determined from sampling the probability distributions in Eq. (\ref{rfromsamples}), and we normalize the projections of the stimulus along each dimension such that they have unit variance.  Note that there is no absolute preference for the sign of the individual stimulus components; rather the spike probability is higher when the two components have opposite signs and lower when they have the same sign.   This is the pattern expected if the neuron in fact is sensitive to the product of the two components, as in the correlation computation of Eq. (\ref{Vcorr}).}
\label{correlator2d}
\end{figure}

To proceed further we sample the probability distributions along the different stimulus dimensions, as explained  in the discussion surrounding Eq. (\ref{rfromsamples}).  Although the reduction from 200 to 4 stimulus dimensions is useful, it still is difficult to examine probability distributions in a four dimensional space.  We proceed in steps, guided by the intuition from the models in Section \ref{models}.  We begin by looking at a two dimensional projection.  We recall  that the optimal estimation strategy involves the correlation of spatial and temporal derivatives.  When we do this [as in Eq's. (\ref{optest-final}) and (\ref{Vcorr})] we find terms involving the product of the time derivative of $s(t)$ with the current value of $c(t)$, as well as the other combination is which $s$ and $c$ are exchanged.  This suggests that we look at the response of H1 in the plane determined by the differentiating filter applied to $s$ and the smoothing filter applied to $c$, and this is shown in Fig. \ref{correlator2d}.  We see the general structure expected if the system is sensitive to a product of the two dimensions:  symmetry across the quadrants, and contours of equal response have an approximately hyperbolic form.  The same structure is seen in the other pair of dimensions, but with the $90^\circ$ rotation expected from the theory.   

\begin{figure}[b]
\centering
\epsfxsize=\linewidth
\epsfbox{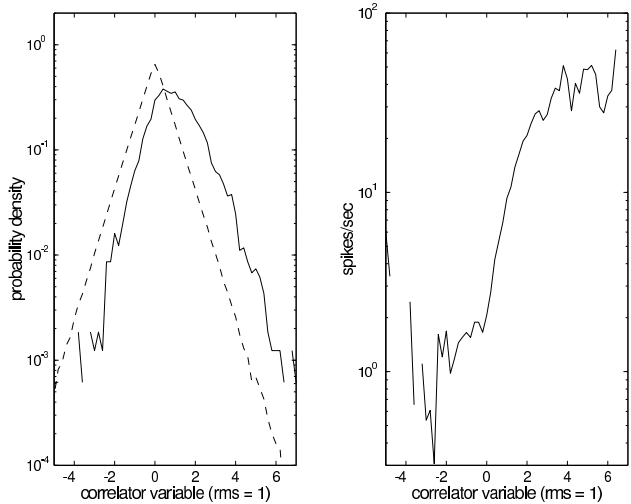}
\caption{Neural response to the correlator variable ${\cal V}_{\rm corr}$ from  Eq. (\ref{Vcorr}).
At left, probability distributions of the correlator variable a priori
(dashed) and conditional on an isolated spike (solid).  Note that since the correlator variable is a nonlinear combination of the four relevant stimulus dimensions, the prior distribution is not Gaussian but in fact almost precisely exponential in form. At right, the spike rate $r(t)$ 
calculated from these distributions using Eq. (\ref{rfromsamples}).}
\label{correlator1d}
\end{figure}

If we take the product structure of the correlator models---and the numerator of the optimal estimation theory prediction in Eq. (\ref{optest-final})---seriously, then we can take the two dimensional projection of Fig. \ref{correlator2d} and collapse it onto a single dimension by forming the product of the two stimulus variables.  We can do the same thing in the other two stimulus dimensions, and then sum these two (nonlinear) stimulus variables to form the  anti--symmetric combination ${\cal V}_{\rm corr}$ from Eq. (\ref{Vcorr}).  The dependence of the spike rate on ${\cal V}_{\rm corr}$ is shown in Fig. \ref{correlator1d}: the rate is modulated by roughly a factor of one hundred in response to changes in this stimulus variable.  We emphasize that this conclusion is derived not by ``manually'' changing the value of the correlator variable, but rather by extracting this nonlinear combination of stimulus variables from the continuous variations of a complex dynamic input.

\begin{figure}[t]
\centering
\epsfxsize=\linewidth
\epsfbox{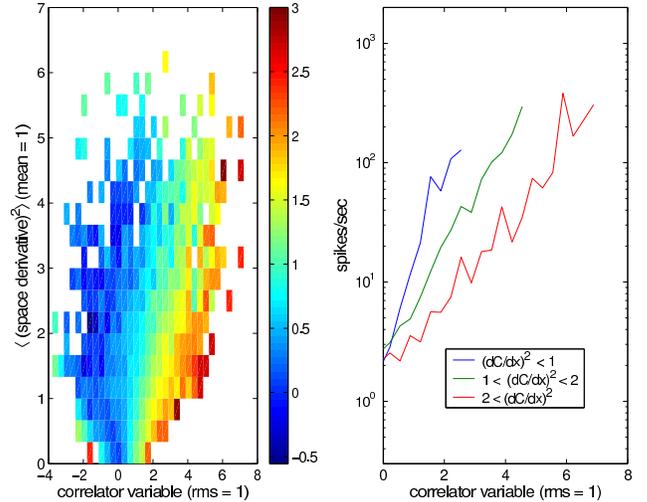}
\caption{Spike probability in two nonlinear dimensions.  At left, the spike rate as a function of the correlator variable ${\cal V}_{\rm corr}$ (as in Fig. \ref{correlator1d}) and the mean square spatial derivative of the movie, as computed from our stimulus projections as proportional to $\cal D$ [Eq. (\ref{D})].  Color scale indicates $\log_{10} r(t)$, with $r$ measured in spikes/sec.  We see a hint that the contours of constant color form a fan, as expected if the neuron responds approximately to the ratio of the correlator variable and the mean square spatial derivative.  At right, we make this clearer by sorting the stimuli into small, medium and large magnitudes of the spatial derivative, effectively taking slices through the two dimensional plot at left and averaging over the distribution of stimuli. The gain of the response to the correlator variable clearly is modulated by the mean square spatial derivative.}
\label{division}
\end{figure}

The correlator variable ${\cal V}_{\rm corr} = s_1\cdot s_3 - s_2 \cdot s_4$ from Eq. (\ref{Vcorr}) is only one of many possible nonlinear combinations of the four stimulus dimensions that emerge from the analysis of $\Delta C$.  It is  predicted by theory to be a central part of the motion computation, but it also defines a quantity that is invariant to a time--independent spatial translation of the visual stimulus; thus the correlator variable automatically incorporates the approximate translation invariance of H1's reponse.  Another such invariant combination is 
\begin{equation}
{\cal D} = s_1^2 + s_2^2 .
\label{D}
\end{equation}
In the predictions of optimal estimation theory [cf. Eq. (\ref{optest-final})], this combination arises because it is proportional to the mean square spatial derivative of the  (temporally filtered) image on the retina.  In Fig. \ref{division}  we show the firing rate $r(t)$ as a function of the two variables ${\cal V}_{\rm corr}$ and $\cal D$.  By exploring this (nonlinear)  two dimensional space we expose a much larger dynamic range of firing rates than can be seen by projecting onto the correlator variable alone as in Fig. \ref{correlator1d}.  We see in the right panel of Fig. \ref{division} that with $\cal D$ fixed there is little evidence of saturation in the plot of spike rate vs. ${\cal V}_{\rm corr}$,   and  the gain in the neural response to the correlator variable is modulated by the magntiude of $\cal D$.

\begin{figure}[b]
\centering
\epsfxsize=\linewidth
\epsfbox{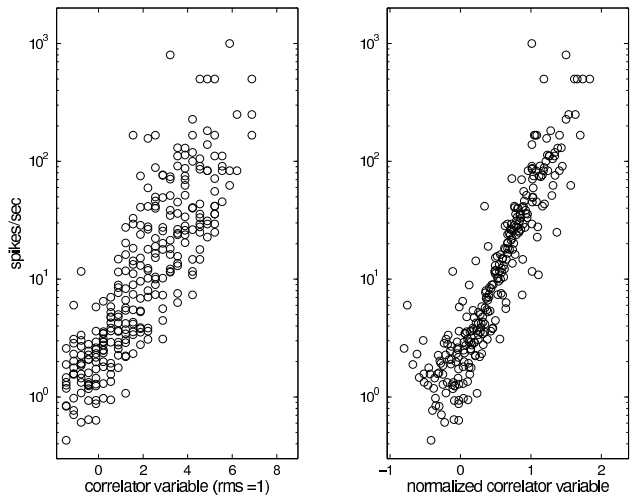}
\caption{Collapsing back to a single stimulus feature.  At left we show the data of Fig. \ref{division} as a scatter plot of spike rate vs. the correlator variable.  Each data point corresponds to a combination of the correlator variable and the mean square spatial derivative; the broad scatter arises from the fact that the spike rate depends significantly on both variables.  At right, we construct a normalized version of the correlator variable corresponding to the form predicted by optimal estimation theory, Eq. (\ref{definesD}); the one arbitrary parameter is chosen to minimize the scatter. Note that the same number of data points appear in both plots; at right many of the points lie on top of one another.}
\label{collapse}
\end{figure}

Another way to look at the results of Fig. \ref{division} is to plot $r(t)$ vs. ${\cal V}_{\rm corr}$ with separate points for the different values of $\cal D$.  Because the rate has a substantial dependence on $\cal D$, there is a large amount of scatter, as seen at the left panel of Fig. \ref{collapse}.  On the other hand, the prediction of optimal estimation theory is that the response should be a function of normalized correlator variable ${\cal V}_{\rm corr}/(B + {\cal D})$ [Eq (\ref{definesD})].  If this prediction is correct, then if we plot $r(t)$ vs this normalized quantity all of the scatter should collapse.  The only uncertainty from optimal estimation theory is the value of the parameter $B$, which depends on detailed assumptions about the statistical structure of the visual world.  We can choose the value of $B$ which minimizes the scatter, however, and the results are shown in the right panel of Fig. \ref{collapse}.
We see  that by constructing the normalized correlator variable predicted from optimal estimation theory we have revealed a dynamic range of nearly $10^3$ in the modulation of  spike rate.  Since the experiment involves only $\sim 8\times 10^3$ isolated spikes, it is difficult to imagine meaningful measurements over a larger dynamic range.

\section{Discussion}

The approach to analysis of neural responses that we have taken here grows out of the  reverse correlation method.  We recall that correlating the input and output of neurons, especially spiking neurons, can be viewed from two very different perspectives, as reviewed in Ref. \cite{rieke+al_97}.  In one view, we imagine writing the firing rate of a neuron as a functional power series in the input signal,
\begin{equation}
r(t) = r_0 + \sum_i K_1(i) {\bf s}_t (i) + {1\over 2} \sum_{ij} K_2 (i,j) {\bf s}_t (i) {\bf s}_t (j) + \cdots ,
\label{series}
\end{equation}
where as in the discussion above ${\bf s}_t (i)$ is the $i^{\rm th}$ component of the stimulus history that leads up to time $t$, and $K_1 , \, K_2 ,\, \cdots$ are a set of ``kernels'' that form the coefficients of the series expansion.
If we choose inputs ${\bf\vec s}_t$ with Gaussian statistics, then by computing the spike--triggered average stimulus we can recover $K_1(i)$, by computing the spike--triggered second moments we can recover $K_2(i,j)$, and so on; this is the Wiener method for analysis of a nonlinear system.  
Poggio and Reichardt \cite{poggio+reichardt_76} emphasized that the correlator model of motion estimation defines a computation that is equivalent precisely to a functional series as in Eq (\ref{series}) but with only the $K_2$ term contributing.  Further, they showed that other visual computations, such as the separation of objects from background via relative motion, can be cast in the same framework, but the minimal nonlinearities are of higher order (e.g., $K_4$ in the case of figure--ground separation). 

Marmarelis and McCann  used the Wiener method to analyze the responses of fly motion sensitive neurons \cite{marmarelis+mccann_73}. Using a pair of independently modulated light spots they verified that $K_1$ makes a negligible contribution to the response, and showed that $K_2$ has the dynamical structure predicted from experiments with double flash stimuli. By construction the second order Wiener kernel describes the same quadratic nonlinearity that is present in the correlator model. Results on the structure of $K_2$ in motion sensitive neurons, as with many other experiments, thus are consistent with the correlator model, but don't really constitute a test of that model.  In particular, such an analysis  cannnot exclude the presence of higher order contributions, such as those described by  Eq (\ref{opt-highsnr1}--\ref{opt-highsnr3}).

Despite its formal generality, the Wiener approach has the problem that it is restricted in practice to low order terms.  If we can measure only  the first few terms in Eq (\ref{series}) we are in effect hoping that neural responses will be only weakly nonlinear functions of the sensory input, and this generally is not  the case.  Similarly, while Poggio and Reichardt  were able to identify the minimum order of nonlinearity required for different visual computations, it is not clear why the brain should use just these minimal terms.  Crucially there is an interpretation of reverse correlation that does not rest on these minimalist or weak nonlinearity assumptions.

In their early work on the auditory system, de Boer and Kuyper  emphasized that if there is a single stage of linear filtering followed by an {\em arbitrary} instantaneous nonlinearity, then with Gaussian inputs the spike--triggered average or ``reverse correlation'' will uncover the linear filter \cite{boer+kuyper_68}.  In our notation, if we can write
\begin{equation}
r(t) = {\bar r}G \left( {\bf \hat v }_1{\bf\cdot }{\bf\vec s}_t \right) ,
\end{equation}
then the spike--triggered average stimulus allows us to recover a vector in stimulus space proportional to the filter or receptive field ${\bf\vec v}_1$ independent of assumptions about the form of the nonlinear function $G$, as long as symmetries of this function do not force the spike--triggered average to be zero.  The hypothesis that neurons are sensitive to a single component of the stimulus clearly is very different from the hypothesis that the neuron responds linearly.  Our approach generalizes this interpretation of reverse correlation to encompass models in which the neural response is driven by multiple stimulus components, but still many fewer  than are required to describe the stimulus completely, as in Eq (\ref{modeldef}).

The success of the receptive field concept as a tool for the qualitative  description of neural responses has led to considerable interest in quantifying the precise form of the receptive fields and their analogs in different systems.  This focus on the linear component of the strongly nonlinear neural response leaves open several important questions.  In the auditory system, for example, observation of spectrotemporal receptive fields with frequency sweep structures does not tell us whether the neuron simply sums the energy in different frequency bands with different delays, or if the neuron has a strong, genuine ``feature detecting'' nonlinearity such that it responds {\em only} when power in one frequency band is followed by power in a neighboring band.  Similarly, the different models of motion estimation discussed in Section  \ref{models} are distinguished not by dramatically different predictions for the spatiotemporal filtering of incoming visual stimuli, but by the way in which these filtered components are combined nonlinearly.  If we hope to explore nonlinear interactions among multiple stimulus components, it is crucial that there not be too many relevant components.  The spike--triggered covariance method as developed here provides us with tools  for counting the number of relevant stimulus dimensions, for identifying these dimensions or features explicitly, and for exploring their interactions.

As far as we know the first consideration of spike--triggered covariance matrices was by Bryant and Segundo  in an analysis of the neural responses to injected currents \cite{bryant+segundo_76}; in many ways this paper was far ahead of its time.  Our own initial work on the spike--triggered (or more generally response--triggered) covariance came from an interest in making models of the full distribution of stimuli conditional on a spike or combination of spikes, from which we could compute the information carried by these events.  In that context the small number of nontrivial eigenvalues in $\Delta C$ meant that we could  make estimates which were more robust against the problems of small sample size \cite{ruyter+bialek_88}.  Roughly ten years later  we realized that this structure implies that the probability of generating a spike must depend on only a low dimensional projection of the stimulus, and that the analysis of spike--triggered covariance matrices thus provides a generalization of reverse correlation to multiple relevant dimensions, as presented here \cite{bialek+ruyter_98}.  

The ideas of the covariance matrix analysis were stated and used in work on adaptation of the neural code in the fly motion sensitive neurons \cite{brenner+al_00}, and in characterizing the computation done by the Hodgkin--Huxley model neuron \cite{aguera+al_03}.  Preliminary results suggest that the same approach via low--dimensional structures may be useful for characterizing the feature selectivity of auditory neurons \cite{sen+al_00}.  In the mammalian visual system the covariance matrix methods have been used in both the retina \cite{schwartz+al_02} and in the primary visual cortex \cite{touryan+al_02, rust_04, rust+al_04} to characterize the neural response beyond the model of a single receptive field.  Most recently this approach has been used to reveal the sensitivity of retinal ganglion cells to multiple dimensions of temporal modulation, and to demonstrate a striking diversity in how these dimensions are combined nonlinearly to determine the neural response \cite{fairhall+al_04}.

In the particular case of motion estimation, the spike--triggered covariance method has made it possible to test important predictions of optimal estimation theory.  The optimal motion estimator exhibits a smooth crossover from correlator--like behavior at low signal to noise ratios to ratio of gradient behavior at high signal to noise ratio \cite{potters+bialek_94}.  In  particular this means that the well known confounding of contrast and velocity in the correlator model should give way to a more ``pure'' velocity sensitivity as the signal to noise ratio is increased.  In practice this means that the contrast dependence of responses in motion sensitive neurons should saturate at high contrast but this saturated response should retain its dependence on velocity.  Further, the form of the saturation should depend  on the mean light intensity and on the statistics of the input movies.  All of these features are observed in the responses of  H1 
\cite{ruyter+al_94},
 but none is  a  ``smoking gun'' for the optimal estimator.  Specifically, the optimal estimator has two very different types of nonlinearity:  the multiplicative nonlinearity of the correlator model, and the divisive nonlinearity of the gradient ratio.  It is this nonlinear structure---and not, for example, a dramatic shift in frequency response or other quasi--linear filtering properties---that seems to be the central prediction of optimal estimation theory.  The spike--triggered covariance matrix method has allowed us to demonstrate directly that both nonlinearities operate simultaneously in shaping the response of H1 to complex, dynamic inputs---the multiplicative nonlinearity is illustrated by Figs \ref{correlator2d} and \ref{correlator1d}, while the divisive nonlinearity is revealed in Figs \ref{division} and \ref{collapse}.  

Although much remains to be done, the demonstration that the nonlinearities in motion computation are of the form predicted from optimal estimation theory helps to close  a circle of ideas which  began with the observation that H1 encodes near--optimal motion estimates, at least under some range of conditions \cite{bialek+al_91, ruyter+bialek_95}.  If the nervous system is to achieve optimal performance (at motion estimation or any other task) then it must carry out certain very specific computations.  Evidence for optimality thus opens a path for theories of neural computation  based on mathematical analysis of the structure of the problem that the system has to solve rather than on assumptions about the internal dynamics of the circuitry that implements the solution.  Except in the very simplest cases, testing these theories requires methods for analysis of the nonlinear structure in the neural response to complex stimuli.

No matter one what one's theoretical prejudices, analyzing neural processing of complex, dynamic inputs requires simplifying hypotheses.  Linearity or weak nonlinearity, central features of the classical Wiener system identification methods, are unlikely to be accurate or sufficient.  The widely used concept of receptive field replaces linearity with a single stimulus template, and much effort has gone into developing methods for finding this single relevant direction in stimulus space.  But already the earliest discussions of receptive fields made clear that there can be more than one relevant stimulus dimension, and that these features can interact nonlinearly.   

The methods developed here  go systematically  beyond the ``single template'' view of receptive fields:  we can count the number of relevant stimulus dimensions, identify these features explicitly, and in favorable cases map the full structure of their nonlinear interactions.  Crucially, essential aspects of the results are clear even from the analysis of relatively small data sets (cf Fig \ref{significance}), so that one can make substantial progress with minutes rather than hours of data.  Further, the possibility of visualizing directly the nonlinear interactions among different stimulus dimensions by sampling the relevant probability distributions allows us to go beyond fitting models to the data; instead one can  be surprised by unexpected features of the neuron's computation, as in recent work on the  retina \cite{fairhall+al_04}.

The idea that neurons are sensitive to low dimensional subspaces within the high dimensional space of natural sensory signals is a hypothesis that needs to be tested more fully.  If correct,   this dimensionality reduction can be sufficiently powerful to render tractable  the otherwise daunting problem of characterizing the neural processing and representation of complex inputs. 

\subsection*{Acknowledgements}
We thank GD Lewen and A Schweitzer for
their help with the experiments.
Discussions with N Brenner and N Tishby were crucial in the
development of our ideas, and we thank also  B Ag\"uera y Arcas, AJ
Doupe, AL Fairhall, R Harris, NC Rust, E Schneidman,  K Sen, T
Sharpee, JA White, and BD Wright for many discussions exploring the application of
these  methods in other contexts.  Early stages of this work were supported by the NEC Research Institute and were presented as part of the summer course on computational neuroscience at the Marine Biological Laboratory; we thank many students and course faculty who asked penetrating questions and had helpful suggestions.  Completion of the work was supported in part by  National Science Foundation Grant IIS--0423039, as part of the program for Collaborative Research in Computational Neuroscience.


{\appendix
\section{Experimental methods}
\label{app-expt}

Spikes from H1 are recorded with a conventional extracellular tungsten
microelectrode (FHC Inc., 3 M$\Omega$), using a silver wire as the
reference electrode. H1 is identified by the combination of its
projection across the midline of the brain and its selectivity to inward
motion \cite{bishop+keehn_66, strausfeld_76, armstrong+al_95}.  Spike arrival times are digitized to 10 $\mu$s precision and stored for further analysis, using a CED 1401plus real time computer. Stimulus patterns are computed using a Digital Signal Processor board (Ariel) based on a Motorola 56001 processor, and consist of frames of nominally 200 vertical lines, written at a frame rate of 500 Hz. Thus the patterns are essentially 1-dimensional, but extended in the vertical direction. They are displayed on a Tektronix 608 monitor (phosphor P31), at a radiance of ${\bar I} = 165 \,{\rm mW}/{\rm  m}^2 \cdot {\rm sr}$. Taking spectral and optical characteristics of the photoreceptor  lens--wave guide into account, this light intensity corresponds to  a flux of  $\sim 4 \times 10^4$ effectively transduced photons/s in each retinal photoreceptor. Frames are
generated in synchrony with the spike timing clock by forcing the DSP to
generate frames triggered by a 2 ms timing  pulse from the CED. Angular dimensions of the display are calibrated using the motion reversal response  of the H1 neuron \cite{gotz_64}.

\end{document}